\documentclass[%
 reprint,
 amsmath,amssymb,
 aps,
 pra,
longbibliography
]{revtex4-2}

\usepackage{graphicx}% Include figure files
\usepackage{dcolumn}% Align table columns on decimal point
\usepackage{bm}% bold math
\usepackage[utf8]{inputenc}
\usepackage[T1]{fontenc}
\usepackage{mathptmx}
\usepackage{etoolbox}
\usepackage{physics}
\usepackage{appendix}
\usepackage{color}
\usepackage{booktabs}

\newcommand{\gs} {5s^2 5p^5\ ^2P^{\circ}_{3/2}}
\newcommand{\gsp}{5s^2 5p^5\ ^2P^{\circ}_{1/2}}

\newcommand{\sixdOne}{5s^2 5p^4(\ ^3P_0)6d\ ^2[2]_{3/2}}
\newcommand{\sixdTwo}{5s^2 5p^4(\ ^3P_1)6d\ ^2[1]_{1/2}}
\newcommand{\sixdThree}{5s^2 5p^4(\ ^3P_1)6d\ ^2[2]_{3/2}}
\newcommand{\sixdFour}{5s^2 5p^4(\ ^3P_1)6d\ ^2[1]_{3/2}}
\newcommand{\sixdFive}{5s^2 5p^4(\ ^1D_2)6d\ ^2[1]_{3/2}}
\newcommand{\sixdSix}{5s^2 5p^4(\ ^1D_2)6d\ ^2[2]_{3/2}}

\newcommand{\sixdEight}{5s^2 5p^4(\ ^3P_1)6d\ ^2[2]_{5/2}}
\newcommand{\sixdNine}{5s^2 5p^4(\ ^3P_0)6d\ ^2[2]_{5/2}}
\newcommand{\sixdTen}{5s^2 5p^4(\ ^3P_2)6d\ ^2[3]_{5/2}}
\newcommand{\sixdEleven}{5s^2 5p^4(\ ^3P_2)6d\ ^2[0]_{1/2}}

\newcommand{\fivedOne}{5s^2 5p^4(\ ^1D_2)5d\ ^2[0]_{1/2}}
\newcommand{\fivedTwo}{5s^2 5p^4(\ ^1S_0)5d\ ^2[2]_{5/2}}
\newcommand{\fivedThree}{5s^2 5p^4(\ ^1S_0)5d\ ^2[2]_{3/2}}

\newcommand{\redmat}[3]{\left \langle #1 \middle|\middle| #2  \middle| \middle| #3 \right \rangle}

\makeatletter
\def\@email#1#2{%
 \endgroup
 \patchcmd{\titleblock@produce}
  {\frontmatter@RRAPformat}
  {\frontmatter@RRAPformat{\produce@RRAP{$\dagger$#1\href{mailto:#2}{#2}}}\frontmatter@RRAPformat}
  {}{}
}%
\makeatother

\graphicspath{
	{figures/}
}

\begin{document}

\title{Control of valence-electron motion in Xe cation using the stimulated-Raman-adiabatic-passage technique}%

\author{Miguel A. Alarc\'on}
\affiliation{Department of Physics, University of Arizona, Tucson, Arizona 85721, USA}

\author{Karl Hauser}
\affiliation{Department of Physics, University of Arizona, Tucson, Arizona 85721, USA}

\author{Nikolay V. Golubev$^\dagger$}
\email{ngolubev@arizona.edu}
\affiliation{Department of Physics, University of Arizona, Tucson, Arizona 85721, USA}

\date{\today}

\begin{abstract}
This work theoretically investigates possibilities of using the Stimulated Raman Adiabatic Passage (STIRAP) and its variants to control a coherent superposition of quantum states. We present a generalization of the so-called fractional STIRAP (f-STIRAP), demonstrating precise control over the mixing ratio of quantum states in the wave packet. In contrast to conventional f-STIRAP, designed to drive a system from an eigenstate into a coherent superposition, our scheme enables arbitrary control over the composition of an already existing superposition state. We demonstrate that an approximate version of this technique---where analytically designed laser pulses with composite envelopes are replaced by simple Gaussian pulses---achieves comparable performance in controlling the dynamics of the wave packet. A limiting case of this scheme, utilizing two pulses with identical Gaussians envelopes and tuned delay and relative phase, is also explored, revealing experimentally accessible pathways for manipulating quantum coherence. We apply our developed techniques to control the ultrafast charge migration in the spin-orbit split ground electronic states of xenon cation via intermediate valence- and core-excited states. Finally, we propose concrete experimental realizations of the developed control schemes in combination with attosecond transient absorption spectroscopy as a method to probe the system.
\end{abstract}

\pacs{Valid PACS appear here}
\maketitle

\section{Introduction}
The coherent control of dynamics in quantum systems has been at the forefront of physics research for decades~\cite{shapiro2012,tannor1985,judson1992,assion1998}. This field has evolved in close connection with the development of laser technologies, especially through the emergence of pulsed and tunable laser sources~\cite{svelto2010}. The invention of ultrafast lasers~\cite{antoine1996,hentschel2001} and their impressive progress~\cite{krausz2009,krausz2016} in recent years have permitted the generation and precise shaping of light with unprecedented temporal and spectral resolution, allowing for real-time observations and manipulations of quantum processes taking place at microscopic scales.

At the quantum level, dynamics arise as a result of coherent superposition of multiple eigenstates of a system. For example, creating a superposition of electronic states, one could trigger the electron motion between different orbitals of an atom or even launch the flow of charge through a molecular chain on subfemtosecond timescales. This phenomenon, commonly referred to as charge migration in the literature~\cite{remacle1998,cederbaum1999}, has become a central topic in ultrafast science, attracting significant interest from both theoretical~\cite{kuleff2005,kuleff2007,lunnemann2008a,kuleff2014,guiotdudoignon2025} and experimental~\cite{weinkauf1994,weinkauf1997,belshaw2012,calegari2014,kraus2015,matselyukh2022} perspectives.

The ability to manipulate charge migration by intense ultrashort laser pulses offers unprecedented opportunities for controlling chemical reactivity and electronic properties of matter~\cite{calegari2014,kraus2015,nisoli2017}. Our understanding of how to initiate, manipulate and guide electronic dynamics has been greatly benefited by numerous studies exploring various aspects of charge migration control, such as the dependence of reaction outcomes on parameters of the ionizing pulse~\cite{weinkauf1994}, effects of spatial orientation of molecules on electron dynamics~\cite{kraus2015}, charge localization by coherent population redistribution between the involved electronic states~\cite{golubev2015,golubev2017qoc}, and the investigation of the effects of electronic decoherence on the ability to steer the nuclear motion in molecules~\cite{golubev2020TGA,fransen2025}. Yet, experimental capabilities to achieve precise control over electronic, and potentially subsequent nuclear, dynamics remain limited, largely due to the stringent requirements for shaping and tuning the laser fields necessary for coherent manipulation of matter at the microscopic level.

Among the various approaches developed for coherent quantum control, the Stimulated Raman Adiabatic Passage (STIRAP)~\cite{gaubatz1990,vitanov2017} technique stands out for its robustness against experimental imperfections and its ability to facilitate complete population transfer without populating intermediate states~\cite{bergmann2015}. This feature is particularly advantageous, since the intermediate states are often prone to get involved in processes such as decoherence or spontaneous decay~\cite{vitanov2017,kral2007,shore2017}, which can severely hinder or even prevent effective control.

Traditional STIRAP involves application of two laser pulses, the ``pump'' and ``Stokes'' pulses, in a counterintuitive sequence to achieve the inversion of populations between two eigenstates of a system via an intermediate state~\cite{shore2017,vitanov2017}. An extended version of STIRAP, designed to transfer population to a coherent superposition of states by manipulating the duration of the Stokes pulse such that it extends to vanish at the same rate as the pump pulse, has been termed fractional (or partial) STIRAP (f-STIRAP)~\cite{marte1991,unanyan1998,vitanov1999,sangouard2004,amniat-talab2005a,amniat-talab2005,chen2007}. STIRAP and its variants have become cornerstone approaches in coherent quantum control, widely used for manipulating various quantum systems including ultracold trapped ions~\cite{sorensen2006}, laser cooled rubidium atoms~\cite{cubel2005}, diatomic molecules~\cite{halfmann1996}, and solid-state platforms such as nitrogen vacancy centers in diamond~\cite{bohm2021}. However, to the best of our knowledge, the use of STIRAP for controlling quantum dynamics, or more generally systems \textit{already present} in a superposition state, remained unexplored until now.

This work aims to provide a theoretical formalism that extends the existing STIRAP protocols to the case that permits control of a quantum superposition state. We report a generalization of f-STIRAP, demonstrating the possibility to arbitrarily manipulate the mixing ratio of quantum states in the wave packet. By taking advantage of the inherent robustness of our approach, we show that an approximate version of generalized f-STIRAP, where the analytically derived laser pulses are replaced by fitted Gaussians, can maintain similar efficacy and drive comparable adiabatic population transfer dynamics. Additionally, we propose an alternative approach, which uses two Gaussian pulses with identical envelopes and optimized phases and delays, achieving a high degree of control. While this scheme deviates substantially from the adiabatic population transfer observed in conventional STIRAP approaches, we demonstrate that key aspects of our method, such as the efficiency and robustness of the scheme with respect to fluctuations of laser pulse parameters, remain unchanged. Furthermore, the proposed technique is readily implementable with existing experimental setups and thus could easily be utilized in realistic applications.

We apply our developed STIRAP methodologies to control charge migration dynamics in the spin-orbit split electronic ground state of the Xe cation. To demonstrate the flexibility of our schemes and bridge the performed simulations with future experimental studies, we numerically simulate the potential implementations of STIRAP with (i) extreme ultraviolet (XUV) sources, utilizing the first excited state of Xe$^+$ as the intermediate state for the population transfer, and (ii) soft x-ray sources, targeting a well-isolated electronic state with a core vacancy. Following the strategy reported in a recent experimental study measuring the electron dynamics in argon cation~\cite{chakraborty2025attosecond}, we propose utilizing attosecond transient absorption spectroscopy (ATAS) to monitor the control over the electron motion in the system. We identify how the population transfer and the presence of the control field affect the absorption spectrum.

The article is organized as follows. Section~\ref{sec:theory} introduces the theoretical basis for STIRAP and f-STIRAP. In Sec.~\ref{sec:Gf-STIRAP}, we generalize the existing f-STIRAP approach to the case when both the initial and final states of the system represent arbitrary coherent superposition states. We first discuss in Sec.~\ref{subsec:Gf-STIRAP_deriv} a straightforward analytic extension of f-STIRAP, and then examine, in Secs.~\ref{subsec:Gf-STIRAP_fitted} and~\ref{subsec:Gf-STIRAP_identicalG}, respectively, the performance of approximate schemes with fitted Gaussian pulses and two Gaussian pulses with identical envelopes controlling their relative delay and phase. Section~\ref{sec:results} contains the results of our numerical simulations demonstrating the applications of our proposed schemes to charge migration control in Xe$^+$. We summarize our results, conclude the paper, and discuss possible directions of future research in Sec.~\ref{sec:conclusions}.

\section{Theoretical Background}
\label{sec:theory}
The methodology and working equations of STIRAP are well known and a detailed analysis of them is presented elsewhere~\cite{shore2017,tannor2007}. Here, we only briefly introduce the basic principles of STIRAP in order to complete our derivations. A typical scenario where STIRAP control is often used is in population transfer in a $\Lambda$-type three-level system described by the following Hamiltonian (we adopt atomic units unless otherwise stated):
\begin{widetext}
\begin{equation}\label{eq:hamil}
    \hat{H}(t) = \begin{pmatrix}
        E_1 & -\Omega_P(t)\cos(\omega_P t) & 0\\[2mm]
        -\Omega_P(t) \cos(\omega_P t) & E_2 & -\Omega_S(t) \cos (\omega_S t +\phi) \\[2mm]
        0 & -\Omega_S(t) \cos (\omega_S t + \phi) & E_3
    \end{pmatrix},
\end{equation}
\end{widetext}
where quantities $E_i$ represent the eigenenergies of the field-free Hamiltonian corresponding to eigenstates $\ket i$, and $\Omega_{P/S}(t)=\mu_{12/23}\mathcal{E}_{P/S}(t)$ denote the time dependent Rabi frequencies for what are commonly called the pump ($P$) and Stokes ($S$) pulses. These pulses are characterized by the corresponding envelopes $\mathcal{E}_{P/S}(t)$, frequencies $\omega_{P/S}$, and relative phase $\phi$ between them, and couple states $\ket{1} \leftrightarrow \ket{2}$ and $\ket{2} \leftrightarrow \ket{3}$ via electric dipole transitions determined by the corresponding matrix elements $\mu_{12}$ and $\mu_{23}$, respectively. The original $P/S$ terminology arose from the consideration of the population transfer along a three-state chain 1-2-3 with the energy ordering $E_1 < E_3 < E_2$, where the pump pulse drives the population ``up'' in energy from state $\ket{1}$ to state $\ket{2}$ and the Stokes pulse moves it ``down'' from state $\ket{2}$ to state $\ket{3}$. The $P/S$ notation becomes meaningless if the system starts in a superposition of states $\ket{1}$ and $\ket{3}$, since depending on the choice of the final state the logic of pump and Stokes pulses could be reversed. Nevertheless, in the followup discussion we stick to this historical notation assuming that the pump and Stokes pulses couple states $\ket{1}$ and $\ket{2}$, and $\ket{2}$ and $\ket{3}$, respectively, independent of the direction of the population transfer.

In order to describe the dynamics and the population transfer in the system described by Hamiltonian~(\ref{eq:hamil}), we solve the time-dependent Schr\"odinger equation (TDSE)
\begin{equation}\label{eq:tdse}
    i \frac{d}{dt}\ket{\psi(t)} = \hat{H}(t) \ket{\psi(t)}.
\end{equation}
Introducing the unitary transformation defined by the operator
\begin{equation}
    \hat{U}(t) = \begin{pmatrix}
        e^{-i E_2 t} e^{i \omega_P t} & 0 & 0 \\
        0 & e^{-i E_2 t} & 0 \\
        0 & 0 & e^{-i E_2 t} e^{i \omega_S t} \\
    \end{pmatrix},
\end{equation}
and switching to the rotating frame it creates, i.e. $\ket {\psi'} = U(t)\ket{\psi(t)}$, we can transform the system to be driven by the following effective Hamiltonian (see, e.g., Ref.~\cite{tannor2007}) 
\begin{equation}
\label{eq:STIRAP_ham}
    \hat{H}'(t) =-\frac{1}{2} \begin{pmatrix}
        -2\delta_{12} & \Omega_P(t) & 0 \\[2mm]
        \Omega_P(t) & 0 & \Omega_S(t)e^{i\phi} \\[2mm]
        0 & \Omega_S(t) e^{-i\phi} & -2 \delta_{23}
    \end{pmatrix}.
\end{equation}
Here, we in addition utilized the rotating wave approximation (RWA) and introduced the so-called one photon detunings $\delta_{12}=E_1-E_2+\omega_P$ and $\delta_{23}=E_3-E_2+\omega_S$, which should be small enough for the RWA to be valid. To simplify the followup derivations, and maximize the efficiency of the population transfer, we assume the two-photon resonance condition, i.e. $\delta=\delta_{23}=\delta_{12}$.

Diagonalizing the Hamiltonian in Eq.~(\ref{eq:STIRAP_ham}) and defining $\Omega(t)=\sqrt{\Omega_P(t)^2 + \Omega_S(t)^2}$, $\Theta(t) = \arctan[\Omega_P(t)/\Omega_S(t)]$, and $\Phi(t) = -\arctan(\Omega(t)/\delta)/2$, we can express the instantaneous eigenenergies and eigenstates of the system succinctly by
\begin{widetext}
%\begin{equation}
\begin{alignat}{3}
    E_0(t) = \delta & \qquad \qquad \ket{\psi_0(t)}&=&\cos \Theta(t) \ket{1} - e^{i\phi} \sin \Theta(t)  \ket{3}, \label{eq:psi0} \\
    E_+(t) = \frac{\delta}{2} \left(1+\sqrt{1+\Omega(t)^2/\delta^2}\right) & \qquad \qquad \ket{\psi_+(t)}&=&\cos\Phi(t)
        \left(e^{-i\phi}\sin \Theta(t) \ket{1} + \cos \Theta(t) \ket{3}\right) + \sin \Phi(t) \ket{2}, \label{eq:psip} \\
    E_-(t) = \frac{\delta}{2} \left(1-\sqrt{1+\Omega(t)^2/\delta^2}\right) & \qquad \qquad  \ket{\psi_-(t)}&=&\sin\Phi(t)
        \left(e^{-i\phi}\sin \Theta(t) \ket{1} + \cos \Theta(t) \ket{3}\right) - \cos \Phi(t) \ket{2}. \label{eq:psim}
\end{alignat}
%\end{equation}
\end{widetext}
Representing the field-driven three-level system in this way allows us to identify ways in which the Rabi frequencies can be modified in order to drive a specific population transfer between the field-free energy levels. In the next section we investigate different scenarios where the initial and final states are superpositions of $\ket 1$ and $\ket 3$, and discuss how to design pulses in order to use STIRAP to drive a desired population transfer.

\section{Generalized f-STIRAP}
\label{sec:Gf-STIRAP}

\subsection{Analytic derivation}
\label{subsec:Gf-STIRAP_deriv}
One of the objectives behind the original proposal of STIRAP was to find a procedure in which one could do population transfer between states $\ket{1}$ and $\ket 3$ without involving state $\ket 2$ since, in a real system, any transient population in this state could result in a loss of coherence or undesired decay. Since state $\ket{\psi_0(t)}$ in Eq.~(\ref{eq:psi0}) is only composed of states $\ket 1$ and $\ket 3$, achieving such a populations transfer is possible if the wave function of the system is made to coincide with $\ket{\psi_0(t)}$ at all times. The fact that this can be achieved by using specific shapes for the Rabi frequencies is the idea at the core of STIRAP and f-STIRAP. 

\begin{figure}
    \includegraphics[width=1.0\linewidth]{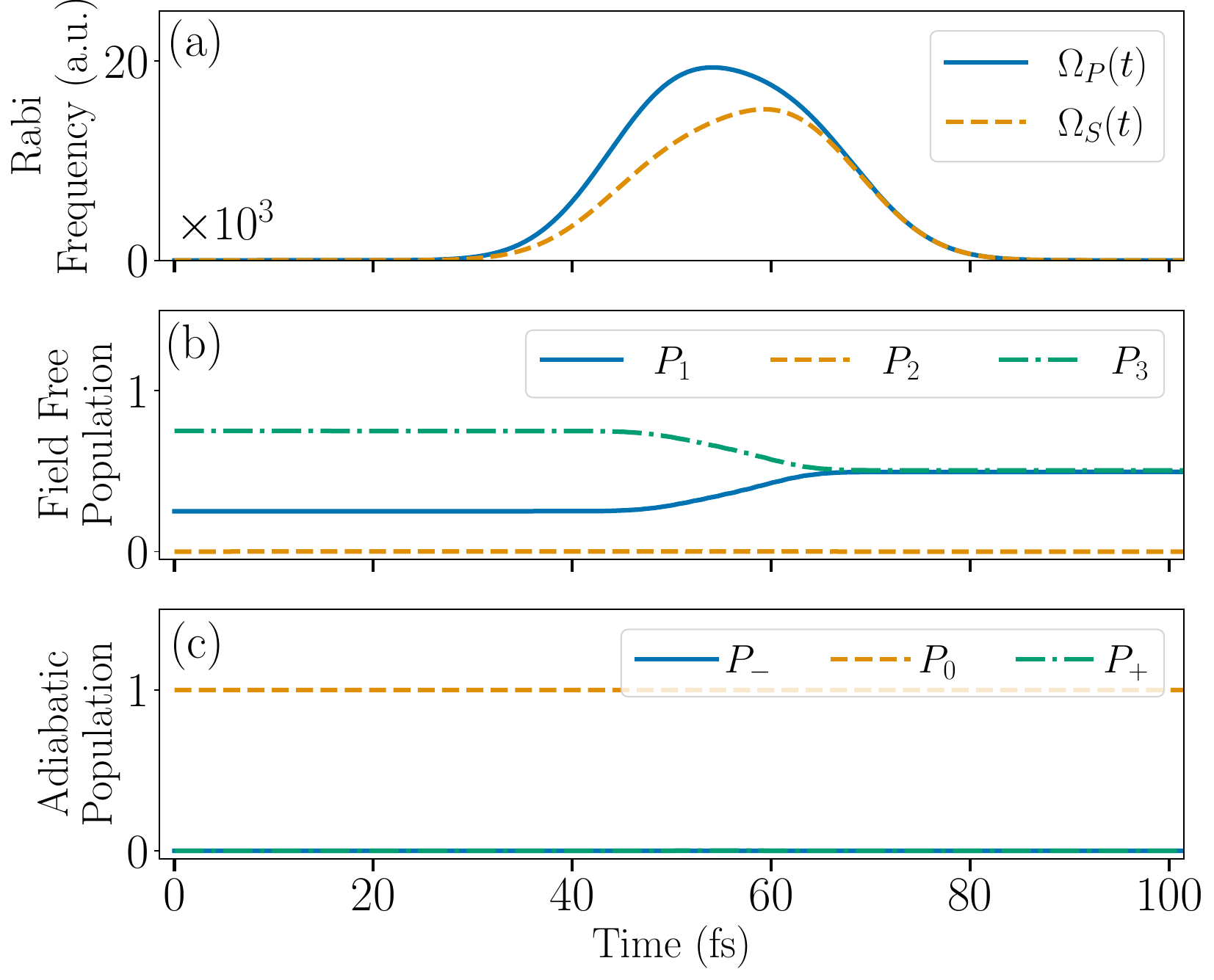}
\caption{Example of the generalized f-STIRAP in a three-level system where a coherent mixture of two lowest eigenstates is controlled by the population transfer via an intermediate state. (a) time dependences of the Rabi frequencies for the pump $\Omega_P(t)$ and Stokes $\Omega_S(t)$ pulses, (b) populations on the field free states, and (c) populations of the adiabatic states in the rotating frame. It is seen that the system is driven from a coherent superposition of states $\ket{1}$ (25\%) and $\ket{3}$ (75\%) to an equal superposition of those states.}
\label{fig:STIRAP}
\end{figure}

In order to obtain a smooth and robust transition between two different superpositions, say transform superposition
\begin{equation}
\label{eq:WF_init}
    \ket{\psi(-\infty)} = \cos \alpha \ket 1 + \sin \alpha \ket 3
\end{equation}
to
\begin{equation}
\label{eq:WF_final}
    \ket{\psi(+\infty)} = \cos \beta \ket 1 + \sin \beta \ket 3,
\end{equation}
the conditions for the laser pulses that induce such a population transfer must be
\begin{equation}
\label{eq:STIRAP_conditions}
    \lim_{t\to -\infty} -\frac{\Omega_P(t)}{\Omega_S(t)} = \tan \alpha \quad \text{and} \quad \lim_{t\to +\infty} -\frac{\Omega_P(t)}{\Omega_S(t)} = \tan \beta.
\end{equation}
These expressions are obtained by directly comparing the initial and the desired final states, Eqs.~(\ref{eq:WF_init}) and~(\ref{eq:WF_final}), with the adiabatic states given by Eqs.~(\ref{eq:psi0})-(\ref{eq:psim}).

At the first glance, the required laser fields can, in principle, be chosen arbitrarily, provided they satisfy the conditions given in Eqs.~(\ref{eq:STIRAP_conditions}). However, one has to remember that the working equations used in the presented derivation were obtained within the RWA and thus the pulses are expected to be nearly resonant with the corresponding transitions and have sufficient duration. Furthermore, the pump and Stokes pulses must have sufficient intensity for the transition between the initial and final states to be slow and smooth so that the quantum state follows the adiabatic state $\ket{\psi_0(t)}$ throughout the interaction thus avoiding mixing state $\ket 2$ into the wave packet.

In the case of conventional STIRAP, i.e. the population inversion with $\alpha=0$ and $\beta=\pi/2$ in Eqs.~(\ref{eq:WF_init}) and~(\ref{eq:WF_final}), $\Omega_S(t)$ has to arrive before $\Omega_P(t)$. This is often referred to as a counterintuitive pulse ordering since it goes against a natural mechanism of the population transfer. Indeed, one would think that from state $\ket 1$ one could drive population into state $\ket 2$ by the action of $\Omega_P(t)$, and then transfer it to state $\ket 3$ with $\Omega_S(t)$. Although this more intuitive ordering allows for a successful population transfer, the true elegance of STIRAP with its reversed order of the pump and Stokes pulses lies in an adiabatic transition mechanism that makes it possible to completely circumvent populating the intermediate state $\ket 2$.

In the case of generalized f-STIRAP, to transfer between arbitrary linear superpositions of $\ket 1$ and $\ket 3$, the asymptotic ratios of the pulse envelopes in Eqs.~(\ref{eq:WF_init}) and~(\ref{eq:WF_final}) must be finite for both positive and negative infinities. This cannot be achieved with two isolated Gaussian pulses since the necessary condition for such a transition is the asymptotic temporal overlap of pump and Stokes laser fields. A possible choice for such pulses could be a combination of Gaussians in the following form:
\begin{equation}\label{eq:two_ps}
\begin{aligned}
    \Omega_P(t) &= E_P \mu_{12} \left[e^{-\left(\frac{t-t_L}{\gamma_L}\right)^2}\sin \alpha + e^{-\left(\frac{t-t_R}{\gamma_R}\right)^2} \sin \beta \right], \\
    \Omega_S(t) &= E_S \mu_{23} \left[e^{-\left(\frac{t-t_L}{\gamma_L}\right)^2}\cos \alpha + e^{-\left(\frac{t-t_R}{\gamma_R}\right)^2} \cos \beta \right],
\end{aligned}
\end{equation}
where $E_{P}$ and $E_{S}$ denote the field strengths of the pump and Stokes pulses, respectively, $t_{L/R}$ are the arrival times and $\gamma_{L/R}$ are widths of the left ($L$) and right ($R$) Gaussians forming the fields.

Let us now illustrate the above described generalized f-STIRAP approach with a concrete analytical and numerical example. Suppose that the system is composed from the energy levels $E_1=0$~eV, $E_2=11.267$~eV, and $E_3=1.306$~eV, coupled via dipole transitions such that $\mu_{12}=\mu_{23}=1$~a.u. and all other dipole moments are zero. These energy levels and couplings between them mimic the spin-orbit split electronic ground state of Xe$^{+}$ coupled via the first excited electronic state (see Sec.~\ref{subsec:Xe_cation} of this paper for more detailed description of the Xe cation). Setting the initial superposition with mixing angle $\alpha=\pi/3$, we would like to generate laser fields that will drive the evolution of the system to a superposition with mixing angle $\beta=\pi/4$. Choosing the time delay, $t_R-t_L = 12.44$~fs, between the left and right Gaussians with identical widths $\gamma_L=\gamma_R=12$~fs, and setting $E_P=E_S=10.88$~TW/cm$^{2}$, we can obtain via Eqs.~(\ref{eq:two_ps}) the required Rabi frequencies of the pump $\Omega_P(t)$ and Stokes $\Omega_S(t)$ fields.

The Rabi frequencies corresponding to the above parameters are shown in Fig.~\ref{fig:STIRAP}(a). We see that the synthesized envelopes are smooth but asymmetric and different from each other. In contrast to conventional STIRAP, where the Stokes pulse precedes the pump pulse, or standard f-STIRAP, where pulses arrive at different times but vanish simultaneously, here we see that both pulses appear and vanish at the same time. Substituting the envelopes to Hamiltonian~(\ref{eq:hamil}) and assuming the field frequencies are tuned exactly on resonances with the corresponding transitions, such that $\omega_P=E_2-E_1$ and $\omega_S=E_2-E_3$, we solve numerically the TDSE, Eq.~(\ref{eq:tdse}), i.e. we calculate the wavefunction $\ket{\psi(t)}$ before introducing the RWA or other approximations. The obtained populations $P_i(t)=|\bra{i}\ket{\psi(t)}|^2$ of the field free states are depicted in Fig.~\ref{fig:STIRAP}~(b). As shown in the figure, the system evolves smoothly from the initial superposition of states $\ket{1}$ and $\ket{3}$ to the desired superposition, while maintaining minimal population in the intermediate state $\ket{2}$. We also compute populations of the adiabatic states, defined by Eqs.~(\ref{eq:psi0})-(\ref{eq:psim}), which are shown in Fig.~\ref{fig:STIRAP}(c). As it was expected, the wavefunction $\ket{\psi(t)}$ matches the time-dependent adiabatic state $\ket{\psi_0(t)}$ perfectly. We, therefore, establish a simple and efficient way to control the mixing ratio of states $\ket{1}$ and $\ket{3}$ in a three-level system using the generalized version of the f-STIRAP scheme.

\begin{figure}
    \centering
    \includegraphics[width=1.0\linewidth]{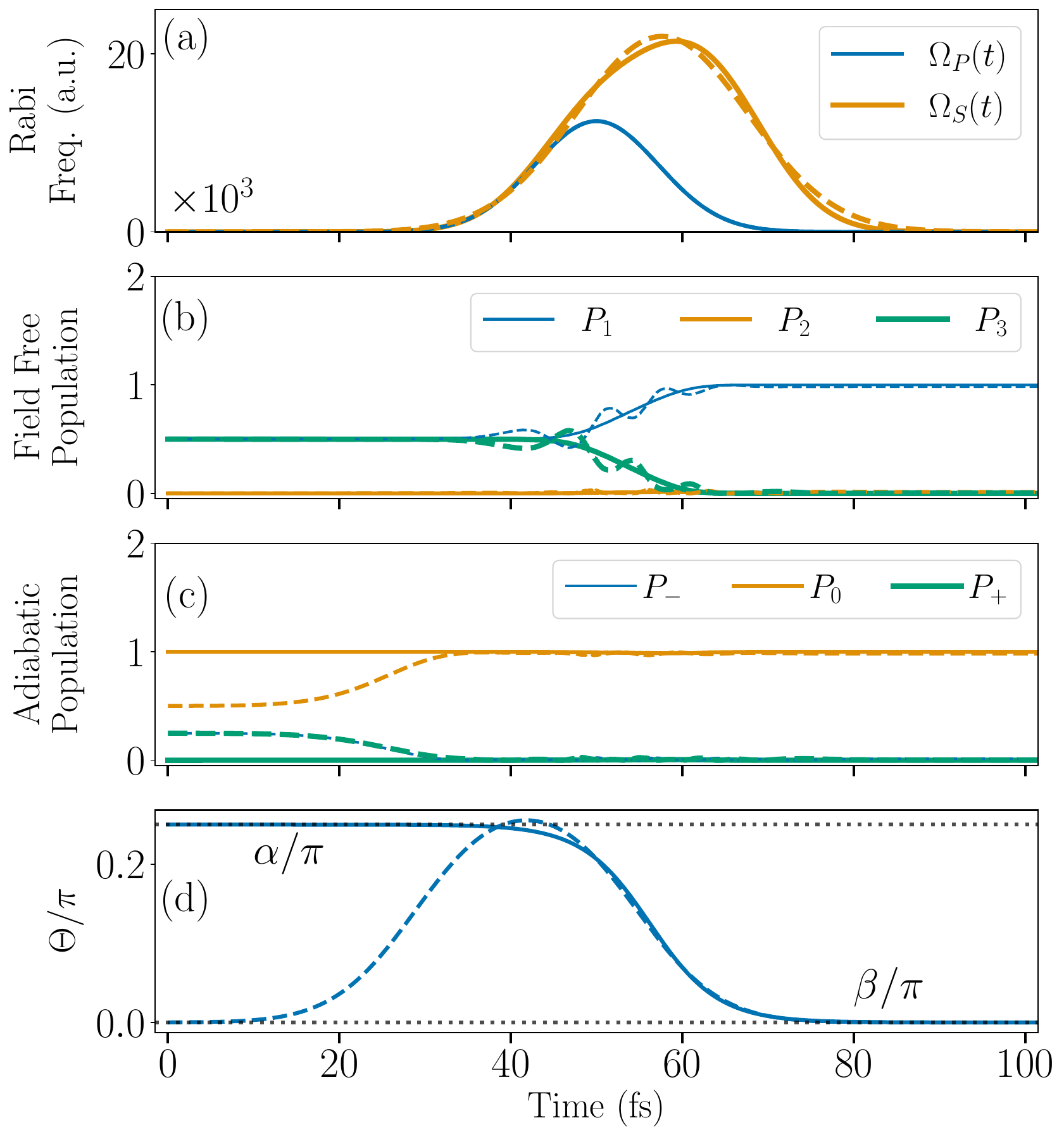}
\caption{Comparison of the analytic generalized f-STIRAP (solid lines) and its approximate version with fitted Gaussian pulses (dashed lines) in controlling a coherent mixture of the two lowest eigenstates via an intermediate state in a three-level system. (a) Rabi frequencies for the pump $\Omega_P(t)$ and Stokes $\Omega_S(t)$ pulses, (b) populations on the field free states, (c) populations of the adiabatic states in the rotating frame, and (d) angle $\Theta(t) = \arctan[\Omega_P(t)/\Omega_S(t)]$ quantifying the difference between the pump and Stokes pulses. It is seen that the system, driven from an equal superposition of states to a single eigenstate, is efficiently controlled by both the analytic and fitted versions of the applied laser fields.}
\label{fig:APPROX}
\end{figure}

\subsection{Generalized f-STIRAP with fitted Gaussians}
\label{subsec:Gf-STIRAP_fitted}
While the analytical expressions derived in Sec.~\ref{subsec:Gf-STIRAP_deriv} enable complete and precise population transfer, the corresponding laser pulses may be challenging to generate in a realistic experimental setting. This is particularly true in the ultrafast regime required to control the electron dynamics, where accurately timing and shaping such unconventional pulses remains extremely challenging with current technologies. An alternative strategy we explore in this section is to approximate the envelope of each pulse by a single Gaussian function. Although such a procedure does not strictly satisfy all the requirements, e.g. those outlined in Eqs.~(\ref{eq:STIRAP_conditions}), the resulting population transfer can nonetheless remain highly efficient due to the intrinsic robustness of STIRAP.

To test the viability of using Gaussian pulses to implement the generalized f-STIRAP, we perform a least square fit to the analytic pulse shapes using the amplitudes $E_{P/S}$, durations $\gamma_{P/S}$, and arrival times $t_{P/S}$ of the Gaussians as the free parameters. This procedure leads, in general, to a new set of parameters that are not necessarily related to those present in Eqs.~(\ref{eq:two_ps}), as these new values include the initial $\alpha$ and final $\beta$ mixing angles. We consider the same system investigated in Sec.~\ref{subsec:Gf-STIRAP_deriv} with same parameters of the left and right fields but explore the population transfer from a superposition with mixing angle $\alpha=\pi/4$ to the eigenstate $\ket{1}$ of the system, i.e. $\beta=0$. For these conditions, the analytically obtained Rabi frequencies (see Eqs.~(\ref{eq:two_ps})) together with the corresponding fitted Gaussians ($E_P=21.76$~TW/cm$^2$, $E_S=67.8$~TW/cm$^2$, $\gamma_P=12$~fs, $\gamma_S=16.68$~fs, $t_P=50$~fs, and $t_S=57.63$~fs) are depicted by solid and dashed lines, respectively, in panel~(a) of Fig.~\ref{fig:APPROX}. As one can see, while in this setting the analytic pump pulse is a perfect Gaussian (since $\beta=0$), the derived Stokes pulse deviates from the pure Gaussian form.

We solve the TDSE with both analytic and fitted Rabi frequencies. The obtained populations of the field free states are depicted by solid (analytic pulses) and dashed (fitted Gaussians) lines in Fig.~\ref{fig:APPROX}(b). We observe that the population transfer dynamics are very similar to each other, although fitted Gaussian pulses lead to additional oscillations during the interaction with the fields. Importantly, the final state populations remain nearly identical, demonstrating the effectiveness of the approximate generalized f-STIRAP in controlling quantum coherences.

It is interesting to note that the populations of the adiabatic states, shown in Fig.~\ref{fig:APPROX}(c), initially differ quite significantly between the analytic and fitted pulse cases.  However, as time progresses, the evolution closely follows the analytical trajectory, and the adiabatic state $\ket{\psi_0(t)}$ becomes dominant. The small residual contributions of the other adiabatic states to the wave packet give rise to the oscillations in the field-free states mentioned in the previous paragraph. The initial discrepancy between the two approaches has little impact on the overall population transfer, as it occurs when the field intensity is still low and the system is only weakly influenced by the interaction. Another way to illustrate this situation is to consider the angle $\Theta(t)$, depicted in Fig.~\ref{fig:APPROX}(d), which quantifies the required relations between the pump and Stokes pulses. As can be seen, although these angles initially differ between the analytic and fitted pulse cases, they start to evolve similarly over time and eventually become nearly identical. Accordingly, when the field intensity raises, the system's response to the fitted Gaussian pulse does not deviate substantially from that to the analytical pulses. This rather small modification in the response and the final populations speaks to the robustness of the adiabatic path followed in STIRAP, which is still followed even with our generalized method.

\begin{figure}
    \centering
    \includegraphics[width=\linewidth]{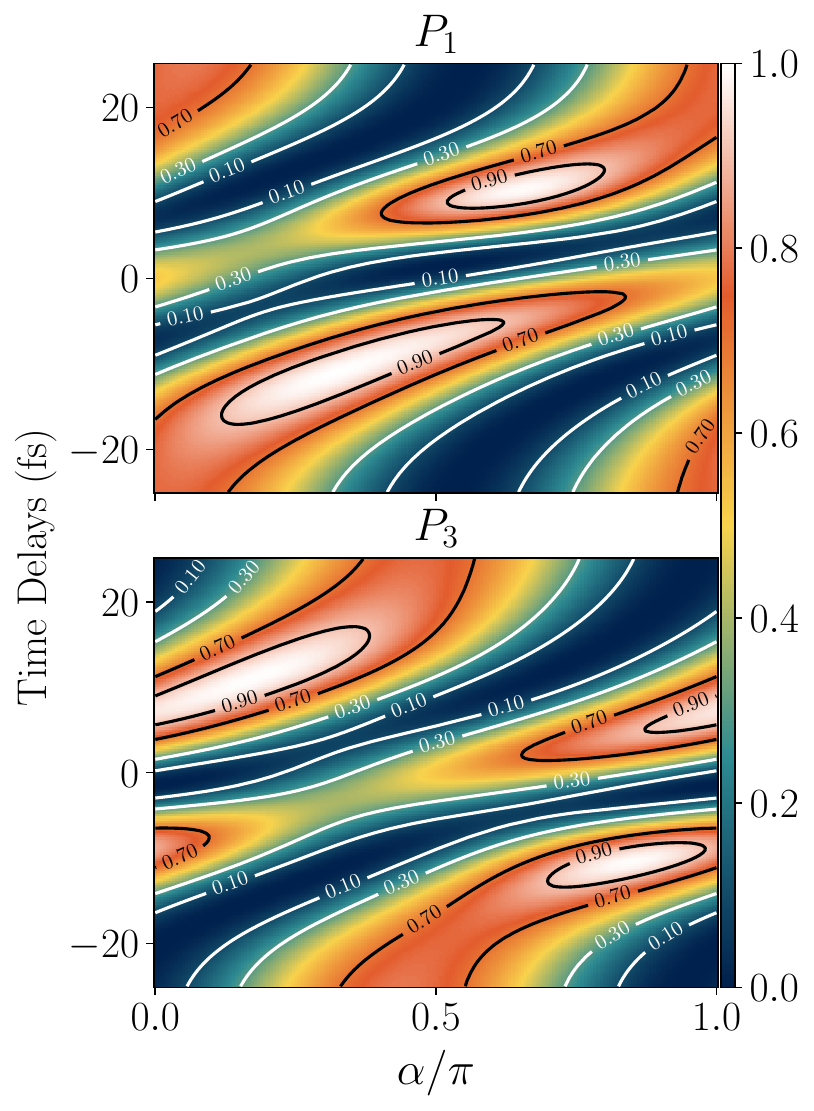}
    \caption{Landscapes showing final populations in states $\ket 1$ (top panel) and $\ket 3$ (bottom panel) evolved from an initial coherent superposition of the same states with various values of mixing angle $\alpha$ (horizontal axis). The population transfer is driven via an intermediate state $\ket 2$ by two Gaussian pulses with identical envelopes, fixed relative phase $\phi = 1.14 \pi$, and controlled relative delays (vertical axis, negative values indicate that the pump pulse precedes the Stokes pulse). It is seen that the system ends up almost entirely in superposition of states $\ket 1$ and $\ket 3$ with only minor amount of population lost in the intermediate state $\ket 2$.}
    \label{fig:delphase}
\end{figure}

\begin{figure}
    \centering
    \includegraphics[width=\linewidth]{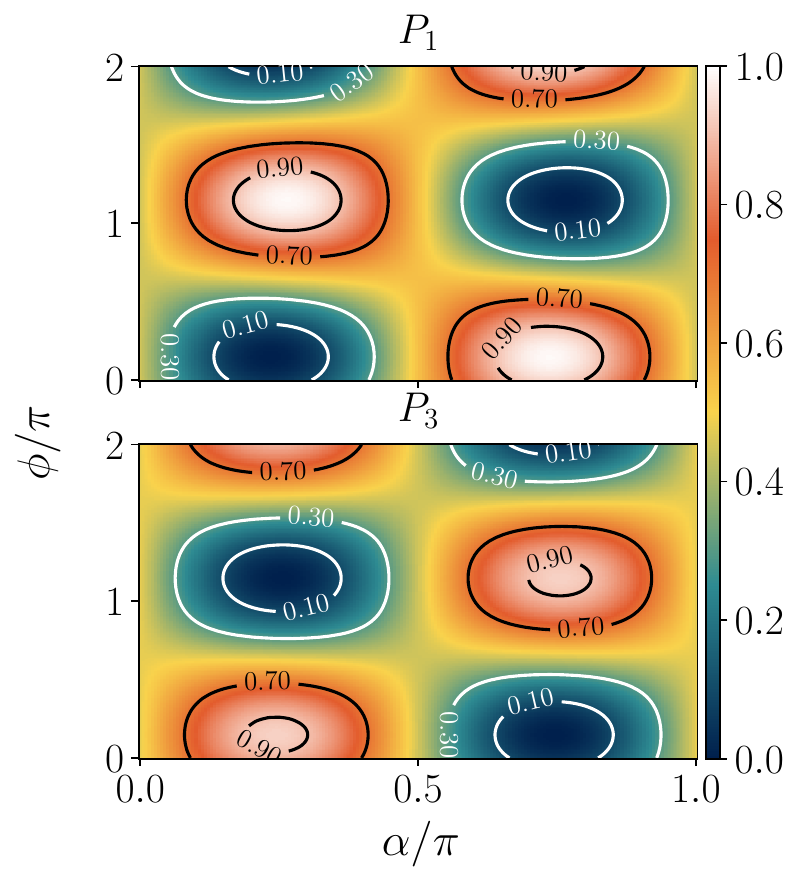}
    \caption{Final populations in states $\ket 1$ and $\ket 3$ for different values of the initial mixing angle $\alpha$ and the relative phase $\phi$ of the pump and Stokes pulses with identical Gaussian envelopes delayed by $12.75$~fs from each other. See the caption of Fig.~\ref{fig:delphase} for a detailed description.}
    \label{fig:initphase}
\end{figure}

\subsection{Population transfer using two pulses with identical Gaussian envelopes}
\label{subsec:Gf-STIRAP_identicalG}
Although the approach with fitted Gaussians demonstrated excellent performance in controlling the coherent wave packet, the resulting pulses were found to differ in their intensities and durations, potentially limiting the energy ranges in which such pulses can be efficiently generated. The present-day experimental capabilities, however, allow for an extremely good control over the time delay and phase offset between two (nearly) identical pulses. In this section, we consider a control scheme with two pulses with identical Gaussian envelopes and tune their phases and the delay between them to achieve the desired population transfer.

In the settings where $\Omega_P(t)$ and $\Omega_S(t)=\Omega_P(t-\Delta t)$ are chosen to be identical to each other up to the delay $\Delta t$, it becomes, in general, not possible to maintain the adiabatic regime of the population transfer since Eqs.~(\ref{eq:STIRAP_conditions}) cannot be satisfied. Therefore, the scheme we investigate technically does not fit to the category of STIRAP anymore. We find that using two pulses with identical Gaussian envelopes with the intensities sufficient to sustain f-STIRAP leads to the majority of the population being transferred to state $\ket{2}$ upon completion of the control sequence, independent of other parameters of the fields. However, by reducing the field intensity, thus taking the evolution of the system away from adiabaticity, we manage to find a rich dependence on pulse parameters that unravels a regime where tuning relative delays and phases allows for the transferal between arbitrary mixing angles.

We performe scans of the laser pulse parameters exploring the final composition of the wave packet starting from various initial conditions. The simulations are performed with pulse amplitudes of $E_P=E_S=3.38$~TW/cm$^2$, widths of $\gamma_L = \gamma_R = 50$~fs, and frequencies tuned on resonances with the corresponding transitions. Figure~\ref{fig:delphase} shows the final population $P_1$ in states $\ket{1}$ (top panel) and $P_3$ in state $\ket{3}$ (bottom panel) as a function of the initial mixing angle $\alpha$ and the time delay $\Delta t$ between two Gaussian pulses with fixed relative phase $\phi = 1.14 \pi$. Complementary, Fig.~\ref{fig:initphase} illustrates the situation when the time delay $\Delta t=-12.75$~fs, i.e. when the pump pulse precedes the Stokes pulse, is fixed and the scan is performed over the relative phase between the pulses. For both of these scans, we see that, in principle, any initial superposition of states can be driven to any desired final mixture of the same states by varying the parameters of the pulses. We highlight that $P_1$ and $P_3$ in both Figs.~\ref{fig:delphase} and~\ref{fig:initphase} add up to one for almost all combinations of the scanned parameters, which indicates that after interacting with the pulses very little population remains in state $\ket 2$. This suggests that for a variety of initial conditions, one can optimize the delay and the relative phase between the two pulses to reach practically any desired final state.

An important characteristic of the investigated control scheme, analogous to that of conventional STIRAP methods, is the smooth dependence of the final state populations on the laser pulse parameters, indicating a notable degree of robustness with respect to small variations of the laser fields. Nonetheless, a major disadvantage of the discussed technique in comparison to STIRAP is that the adiabaticity is sacrificed. We find that the state $\ket 2$ is transiently populated during the evolution of the system. Accordingly, in a realistic application, this intermediate state must be selected such that its transient occupation does not induce decoherence or activate competing processes that could compromise the fidelity of the control.

\begin{figure*}
    \includegraphics[width=\linewidth]{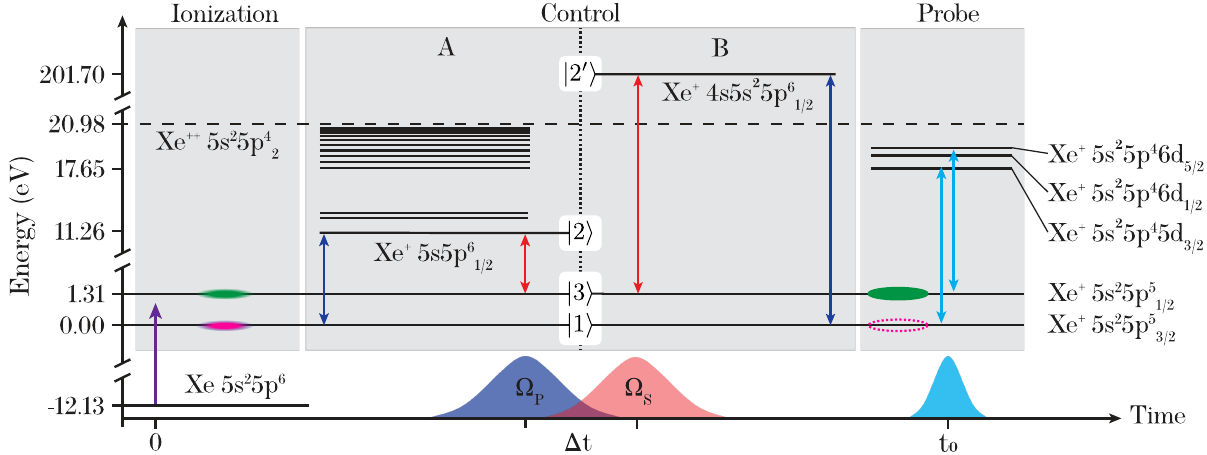}
    \caption{Schematic representation of the relevant energy levels, couplings, and applied laser fields for studying and controlling the charge migration dynamics in Xe$^+$. Following ionization, a linear combination (represented by the ellipsoids) of the two lowest spin-orbit split ionic states is created. Appropriately tuned ``pump'' and ``Stokes'' laser pulses are applied to the system in order to redistribute populations between the initial states using a variant of the STIRAP technique. Depending on the frequency of the utilized control pulses, the intermediate state for the population transfer can either be (A)~one of the low-lying valence electronic states of the cation, or (B)~energetically deep core ionic state. The electron dynamics and the influence of the control fields on the system are quantified by measuring the transmission of weak XUV probe pulses as a function of time delay.}
    \label{fig:Level_Diagram}
\end{figure*}

\section{Control of charge migration in Xe cation}
\label{sec:results}

We now turn to the applications of the developed control schemes for manipulating ultrafast electron dynamics in a realistic atomic system. As a case study, we consider the singly ionized xenon atom and explore the feasibility of controlling the charge migration dynamics within its valence shells. To trace the evolution of the system in time and observe the effects of the control fields, we utilize the renowned ATAS technique~\cite{goulielmakis2010,santra2011,golubev2021} which has been successfully used in the recent past to measure the ultrafast electron dynamics in atomic~\cite{kobayashi2018,chakraborty2025attosecond} and molecular~\cite{matselyukh2022} systems. Accordingly, we implement and simulate the existing schemes for passive observation of the charge migration, and extend them by including the possibility of controlling the electron motion. We attempt to connect, although purely theoretically, the utilized laser pulses with existing experimental techniques suggesting the potential roads for implementation in experiments capable of verifying our predictions. The proposed ionization, control, and probe steps are summarized in Fig.~\ref{fig:Level_Diagram}. We investigate and discuss each of the involved steps in detail below.

\subsection{Description of the system}
\label{subsec:Xe_cation}
In order to perform accurate simulations of the Xe cation and include interactions with realistic control and probe laser fields, it is necessary to consider more than the three states used to define the basic STIRAP Hamiltonian. To access the electronic states of the system, we employ a semiempirical approach based on data from the NIST Atomic Spectra Database~\cite{NIST_ASD_2024}. Given that Xe is such a complex atom and that the states involved in the process lie high in energy, we found that \textit{ab initio} methods are unable to recover state energies comparable to the accurate experimental values reported in NIST. Therefore, in the simulations presented in this study we utilize the electronic states listed in the NIST database, directly importing the state energies and using the corresponding electronic configurations to compute the transition dipole moments as detailed in the Appendix. The application of our control techniques to the xenon cation serves as a proof of concept. Any corrections to the calculated dipole moments would only affect the reported laser intensities, while the allowed couplings and energy levels are accurately captured by this model. Hence, the suggested laser frequencies and pulse durations should serve as useful guidelines for potential experimental implementations.

Since it has been observed that ionization with a strong infrared~(IR) pulse can induce the coherence between lowest electronic states of ionic systems~\cite{goulielmakis2010,chakraborty2025attosecond}, we choose the four-degenerate $5s^2 5p^5 (^2P^{\circ}_{3/2})$ and two-degenerate $5s^2 5p^5 (^2P^{\circ}_{1/2})$ spin-orbit split states of Xe$^+$ as our target states for control. Creating a coherent superposition of these states, one can induce the dynamics of the electron density with the oscillation period of approximately 3.17~fs originating from the energy separation of $1.306$~eV between them~\cite{NIST_ASD_2024}. In terms of the labels used in Secs.~\ref{sec:theory} and~\ref{sec:Gf-STIRAP}, these two states correspond to states $\ket 1$ and $\ket 3$, respectively.

For the intermediate auxiliary state, i.e. state $\ket{2}$, we propose two alternative schemes based on different laser sources that target different electronic states in the ion's spectrum. First, we consider the nearest low-lying bound electronic state $5s5p^6 (^{2}S_{1/2})$ separated from the lower ionic ground state by $11.267$~eV. While this arrangement of states has, in principle, already been investigated in Sec.~\ref{sec:Gf-STIRAP}, the close proximity of other states, in particular the state $5s^25p^4(^3P_2)6s (^2[2]_{5/2})$, which is approximately 272~meV above state $\ket{2}$, imposes additional requirements on the utilized control pulses. In the follow-up applications of the STIRAP techniques to Xe$^+$, we will therefore employ slightly longer pulses with narrower spectral bandwidths to minimize the coupling to undesired nearby electronic states. The laser frequencies required to perform the STIRAP control in this regime belong to the XUV energy ranges and can be efficiently synthesized in a laboratory employing the high-harmonic generation~(HHG) process. The investigation of this scenario is presented in Sec.~\ref{subsec:XUV_control}.

The second option for the STIRAP control we consider in this study is to utilize the state $4s5s^25p^6 (^{2}S_{1/2})$, which we will refer to as $\ket{2'}$ to distinguish it from the low energetic intermediate state $\ket{2}$ discussed before, with a vacancy in the core $4s$ orbital. The energy separation of this state from the ground electronic state of the ion is approximately 201.7~eV~\cite{vaughan1985}, which requires x-ray pulses to implement the control scheme. In principle, core ionic states are surrounded by a large number of neighboring excited states, similar to the dense spectra typically observed in the valence region. However, binding energies of all other inner shell vacancies are energetically well separated from the $4s$ one. This implies that all the one-electron excited states will be far removed from our target intermediate state. All other states within the bandwidth of our control pulses will be at least doubly-excited electronic states, which are not dipole coupled to the spin-orbit split ground states. Using this $4s$ shell vacancy as the intermediate state represents a nearly perfect example of a three-level system, where the accuracy and efficiency of the STIRAP-like schemes are only limited by source constrains on pulse shapes, intensities and frequencies. We discuss the charge migration control with X-ray pulses via the core ionic $\ket{2'}$ state in Sec.~\ref{subsec:Xray_control}.

The ATAS scheme we use for the probe step is, in principle, flexible regarding the choice of final states for the absorption transitions. The only requirement for an absorption line to be sensitive to coherent electron dynamics in the initial states is that these states are coupled to a common final state by the probe pulse. In this work, we target a rather dense spectral region between $16.25$ and $18.25$~eV. Due to the difference in strengths of the involved dipole moments, some lines appear much clearer in the absorption spectra than the others. The most relevant lines are identified in Table~\ref{tab:lines_ths} and will also be marked in the simulated ATAS spectra later in this section.

\begin{table}
\begin{ruledtabular}
\begin{tabular}{cccc} Initial State & Final State & Symbol & Energy (eV) \\
\midrule
$\gs$	&	$\fivedOne$	  &	$\nu^{3/2}_1$	&	16.745426	\\[3mm]
	&	$\fivedTwo$	  &	$\nu^{3/2}_2$	&	16.932476	\\[3mm]
	&	$\fivedThree$ &	$\nu^{3/2}_3$	&	17.11758	\\[3mm]
	&	$\sixdEleven$ &	$\nu^{3/2}_4$	&	17.199905	\\[3mm]
	&	$\sixdTen$	  &	$\nu^{3/2}_5$	&	17.245493	\\[3mm]
	&	$\sixdNine$	  &	$\nu^{3/2}_6$	&	17.901446	\\[3mm]
	&	$\sixdTwo$	  &	$\nu^{3/2}_7$	&	18.005323	\\[3mm]
	&	$\sixdEight$  &	$\nu^{3/2}_8$	&	18.139575	\\[3mm]
	&	$\sixdNine$	  &	$\nu^{3/2}_9$	&	18.216733	\\[1mm]
\midrule
$\gsp$	&	$\sixdOne$	  &	$\nu^{1/2}_1$	&	16.564679	\\[3mm]
	&	$\sixdTwo$	  &	$\nu^{1/2}_2$	&	16.6989	    \\[3mm]
	&	$\sixdThree$  &	$\nu^{1/2}_3$	&	16.787873	\\[3mm]
	&	$\sixdFour$	  &	$\nu^{1/2}_4$	&	17.053801	\\[3mm]
	&	$\sixdFive$	  &	$\nu^{1/2}_5$	&	17.735577	\\[3mm]
	&	$\sixdSix$	  &	$\nu^{1/2}_6$	&	17.862535	\\[1mm]
\end{tabular}
\end{ruledtabular}
\caption{Strongest dipole-allowed electronic transitions from the spin-orbit split ground state of Xe$^{+}$ in the energy range from $16.25$ to $18.25$~eV. The electronic configurations and the energies are taken from the NIST Atomic Spectra Database~\cite{NIST_ASD_2024}.}
\label{tab:lines_ths}
\end{table}

\begin{figure*}
    \includegraphics[width=\linewidth]{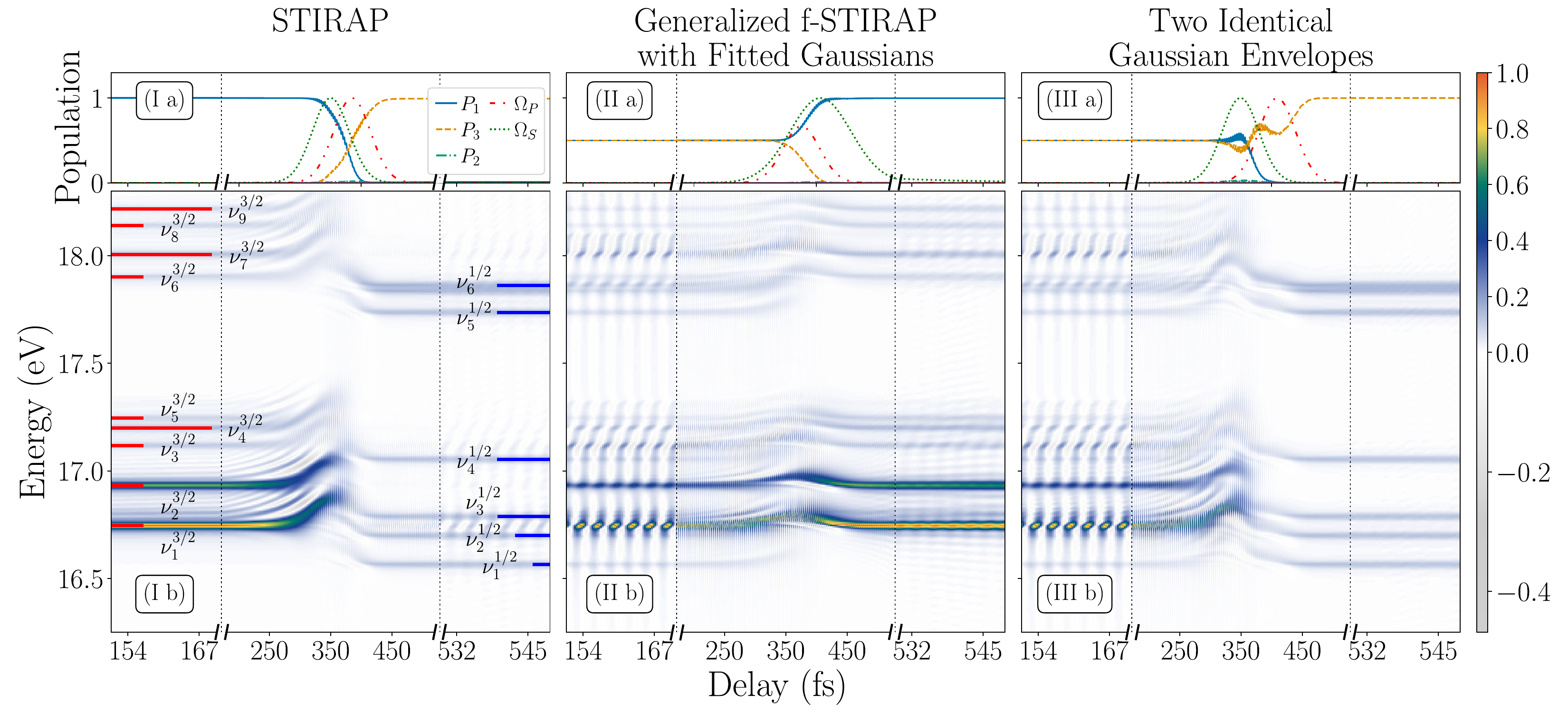}
    \caption{Control of electron motion in spin-orbit split $5s^2 5p^5 (^2P^{\circ}_{3/2})$ and $5s^2 5p^5 (^2P^{\circ}_{1/2})$ ground electronic states of the xenon cation via low-lying $5s5p^6 (^{2}S_{1/2})$ excited electronic state using STIRAP and its variants. Top panels show the applied laser fields and the resulting population transfer, while bottom panels present the corresponding time-resolved transient absorption spectrograms. Left column: Conventional STIRAP scheme demonstrating the inversion of populations between $^2P^{\circ}_{3/2}$ and $^2P^{\circ}_{1/2}$ states. Colored horizontal bars indicate the expected positions of the absorption lines listed in Table~\ref{tab:lines_ths}. Different lengths are used for clarity in the labels. Middle column: Generalized fractional STIRAP with fitted Gaussian pulses designed to drive the coherent wave packet composed of an equal mixture of $^2P^{\circ}_{3/2}$ and $^2P^{\circ}_{1/2}$ states to $^2P^{\circ}_{3/2}$ state. Right column: Scheme with two delayed Gaussian pulses with identical envelopes pushing the same initial wave packet to $^2P^{\circ}_{1/2}$ state. In all three cases, the changes observed in the absorption spectra clearly reflect the achieved control over the electron motion in the system.}
    \label{fig:fractspectrum}
\end{figure*}

The electronic energy levels discussed earlier in this section are those that play a significant role in the dynamics, laser control, and probe steps. However, for completeness, our simulations include all Xe$^+$ energy levels available in the NIST database ($998$ states in total). We simulate the electron dynamics and the action of the control and probe fields on the system by numerically solving the TDSE using the split operator method~\cite{tarana2012,alarcon2023} and assuming the validity of the dipole approximation. All laser pulses used in our calculations, including a hypothetical ionizing pulse, are polarized along the $z$ direction. This allows us to constrain our simulations by considering only the states with a certain value $m$ of the projection of the total angular momentum $J$ on the $z$ axis. Picking $m=1/2$, we further truncate the simulations to a subsystem of $162$ states, thus removing degenerate states not accessible due to selection rules.  We evaluate the time-dependent polarization response and compute the absorption cross-section following the procedure described, e.g., in Ref.~\cite{wu2016}. We, therefore, perform a scan of the dynamics in the system by varying the pump-probe delay $t_0$. The time grid used to propagate the wavefunctions when using XUV pulses spans from 0 to $650$~fs using a $13$~as time step. For x-ray wavelengths, the wave function is propagated from 0 to $300$~fs using a $3$~as time step. To numerically converge the Fourier transform of the computed polarization response, we apply the exponential decay filter with a characteristic timescale of $10$~fs, thus mimicking the finite lifetime of the electronic states reached by the probe pulse and other macroscopic decoherence effects that are unavoidable in a realistic experiment.

\subsection{XUV control via low-lying bound electronic state}
\label{subsec:XUV_control}
When XUV pulses are produced via HHG using a driver with frequency $\omega$, the generated odd harmonics are separated from each other by $2\omega$ frequencies. Since it has been shown~\cite{lai2013,mandal2022} that HHG can be sustained over a wide range of IR driving frequencies, it should be possible to generate a train of harmonics using a driver with a frequency close to half of the spin-orbit splitting between the ground states of the Xe ion. This implies that the spacing between the harmonics will be equal to the spin-orbit splitting, and thus pairs of subsequent harmonics will fulfill the two photon resonance condition which is desirable for STIRAP.

The ideal setup we have in mind uses an IR driver of about 0.653~eV, which makes the $15$th and $17$th harmonics closely in resonance with the $\ket{3}\leftrightarrow\ket{2}$ and $\ket{1}\leftrightarrow\ket{2}$ transitions, respectively. We will utilize these two harmonics with photon energies 9.795 and 11.101~eV, respectively, in our STIRAP control schemes assuming the ability to manipulate their envelopes, relative delay, and phase offset between them whenever necessary. Within this same framework, one can use the $25$th and $27$th harmonics, with energies 16.325 and 17.631~eV, respectively, to probe the electronic dynamics. We assume a probe laser field is composed of these two frequencies with Gaussian envelopes with a 1~fs duration and $8\times 10^9$~W/cm$^2$ intensity.

As an illustrative example, let us first consider the typical case where the system starts in the lowest eigenstate $\ket{1}$. Here we investigate the application of conventional STIRAP to achieve the population transfer to the eigenstate $\ket{3}$. The utilized pump and Stokes laser pulses, depicted in Fig.~\ref{fig:fractspectrum}(a-i), have Gaussian envelopes with the following parameters: $E_P=E_S$ = $0.23$~TW/cm$^2$, $\gamma_P=\gamma_S = 50$~fs, and $\Delta t = -34.8$~fs. The population transfer between the states $\ket{1}$ and $\ket{3}$, together with the transient population of the intermediate state $\ket{2}$, is shown in the same panel of Fig.~\ref{fig:fractspectrum}. As expected, the applied control field causes nearly complete population transfer from state $\ket{1}$ to $\ket{3}$.

The calculated absorption cross-section is presented in Fig.~\ref{fig:fractspectrum}(b-i). It is seen that before the arrival of the control pulses, the absorption spectrum, originating from transitions from state $\ket{1}$, is stationary in time and contains strong absorption lines denoted $\nu^{3/2}_1$ and $\nu^{3/2}_2$. One also finds weaker signatures of other strong transitions to $d$ states in lines $\nu^{3/2}_{3}$ to $\nu^{3/2}_{9}$. As the control pulses come in, the energy positions and profiles of the absorption lines are modified due to the Stark shift induced by the strong lasers~\cite{chakraborty2025attosecond}. Additionally, as the population gets transferred, the spectrum starts containing lines corresponding to the $\nu^{1/2}_{1}$ to $\nu^{1/2}_6$ transitions taking place from state $\ket{3}$. These transitions are weaker compare to those from state $\ket{1}$ and have more uniform strengths. One can also observe that as the pulses have passed, some of the absorption lines start weakly oscillating. This evidence of an unintentionally created coherence is due to the imperfect population transfer that does not achieve $100\%$ efficiency. The latter takes place due to the fact that the utilized STIRAP pulses are slightly detuned from the corresponding resonances, which causes a small amount of population to remain present in state~$\ket{1}$.

We investigated the performance of various versions of our generalized f-STIRAP schemes. In Figs.~\ref{fig:fractspectrum}(a-ii) and \ref{fig:fractspectrum}(b-ii), we show the dynamics of the population transfer and the corresponding absorption spectra, respectively, for the charge migration control with two fitted Gaussian laser pulses. Following the logic discussed in Secs.~\ref{subsec:Gf-STIRAP_deriv} and~\ref{subsec:Gf-STIRAP_fitted}, we first generated analytic laser pulses, aiming to perform population transfer from an equal superposition of states $\ket{1}$ and $\ket{3}$ to eigenstate $\ket{1}$, and then fit the obtained waveforms with Gaussian envelopes. The obtained parameters are $E_P = 0.00501$~TW/cm$^2$, $\gamma_P = 50$~fs, $E_S = 0.113$~TW/cm$^2$, $\gamma_S = 82.90$~fs, and $\Delta t = 31.56$~fs. It is seen that the system, initially present in a superposition state, demonstrates perfect charge migration dynamics which is reflected in the oscillations of the absorption lines. The control field brings the system to the eigenstate $\ket{1}$ thus terminating the electron motion. Accordingly, after the action of the control field the absorption lines become nearly stationary with small oscillations still present due to a residual amount of populations remaining in state $\ket{3}$.

We also implement the control scheme with two delayed pulses with identical Gaussians envelopes. We fix the intensities and pulse durations to be $E_P = E_S = 0.273$~TW/cm$^2$ and $\gamma_P = \gamma_S = 50$~fs, respectively, phase offset $\phi = 0$, and optimize the delay, $\Delta t = -60$~fs, between the pulses such that an equal superposition of the initial states $\ket{1}$ and $\ket{3}$ will be transformed to the eigenstate $\ket{3}$ of the system. In order to obtain this optimized transition, the frequency of the pump has been slightly modified to $11.02$~eV, and the Stokes pulse has also been adjusted accordingly to keep the two-photon resonance condition valid. The results, shown in Figs.~\ref{fig:fractspectrum}(a-iii) and~\ref{fig:fractspectrum}(b-iii), demonstrate the dynamics in the system driven by such pulses. As one can see, we were able to achieve nearly perfect population transfer, although in this case the intermediate state $\ket{2}$ gets transiently populated during the action of the control field.

To speak about the feasibility of our approach and discuss possible experimental realizations of the proposed control schemes, we note two experimental studies that have shown the capability of implementing HHG with driving pulses very close to the desired situation we investigated above. Using an approximately 2000~nm driving pulse it was possible to generate low XUV harmonics using a solid state target~\cite{peterka2023} and a gas system~\cite{mahieu2012}. This suggests that with a powerful enough driving laser capable of generating a large enough cross section for HHG and an appropriate choice of the target system, it would be possible to obtain the laser frequencies we utilized for the control.

The second challenge in generating HHG pulses with the desired characteristics is the selective isolation and modification of the two relevant harmonics. To tackle this, we propose using a time-delay compensating monochromator~(TDCM)~\cite{poletto2009,li2024,braig2022}. Such a device uses a pair of diffracting gratings to separate the colors in the pulse train, and it uses the multiple bursts generated in the train, to elongate the pulse envelope. This would in principle make it possible to separate the two harmonics required for the control scheme and time them appropriately.

An alternative approach for the direct manipulation of the HHG pulses is a setup which instead of using a single source, uses two. Tuning properties of the IR drivers, one can achieve precise control over parameters of the generated harmonics. Furthermore, this allows for an additional degree of control over the generated fields and can be used to adjust, e.g., light polarization. By applying polarized light to the system, it is possible to explore more sophisticated pathways in the population transfer and thus to achieve more precise control over the dynamics. An example would be as follows: assume the source with $\Omega_P(t)$ (the 17th harmonic) has $\sigma_{+/-}$ polarization while $\Omega_S(t)$ (the 15th harmonic) is linearly polarized in the $z$ direction. With these pulses, the three-state system is modified such that $\ket{1}=\ket{5s^2 5p^5 (^2P^{\circ}_{3/2}),m=3/2}$, $\ket{2}=\ket{5s5p^6 (^{2}S_{1/2}),m=1/2}$ and $\ket{3}=\ket{5s^2 5p^5 (^2P^{\circ}_{1/2}),m=1/2}$. In these settings, the transitions induced by the pump and Stokes pulses become completely isolated since, for example, $\Omega_P(t)$ cannot cause the undesired coupling between states $\ket{3}$ and $\ket{2}$. Therefore, the true Hamiltonian of the system becomes much closer to the ideal form of Eq.~(\ref{eq:hamil}). With this, one can achieve subfemtosecond laser control of the charge migration which could be crucial in systems where the decoherence or decay processes need to be taken into account.

\begin{figure}
    \centering
    \includegraphics[width=\linewidth]{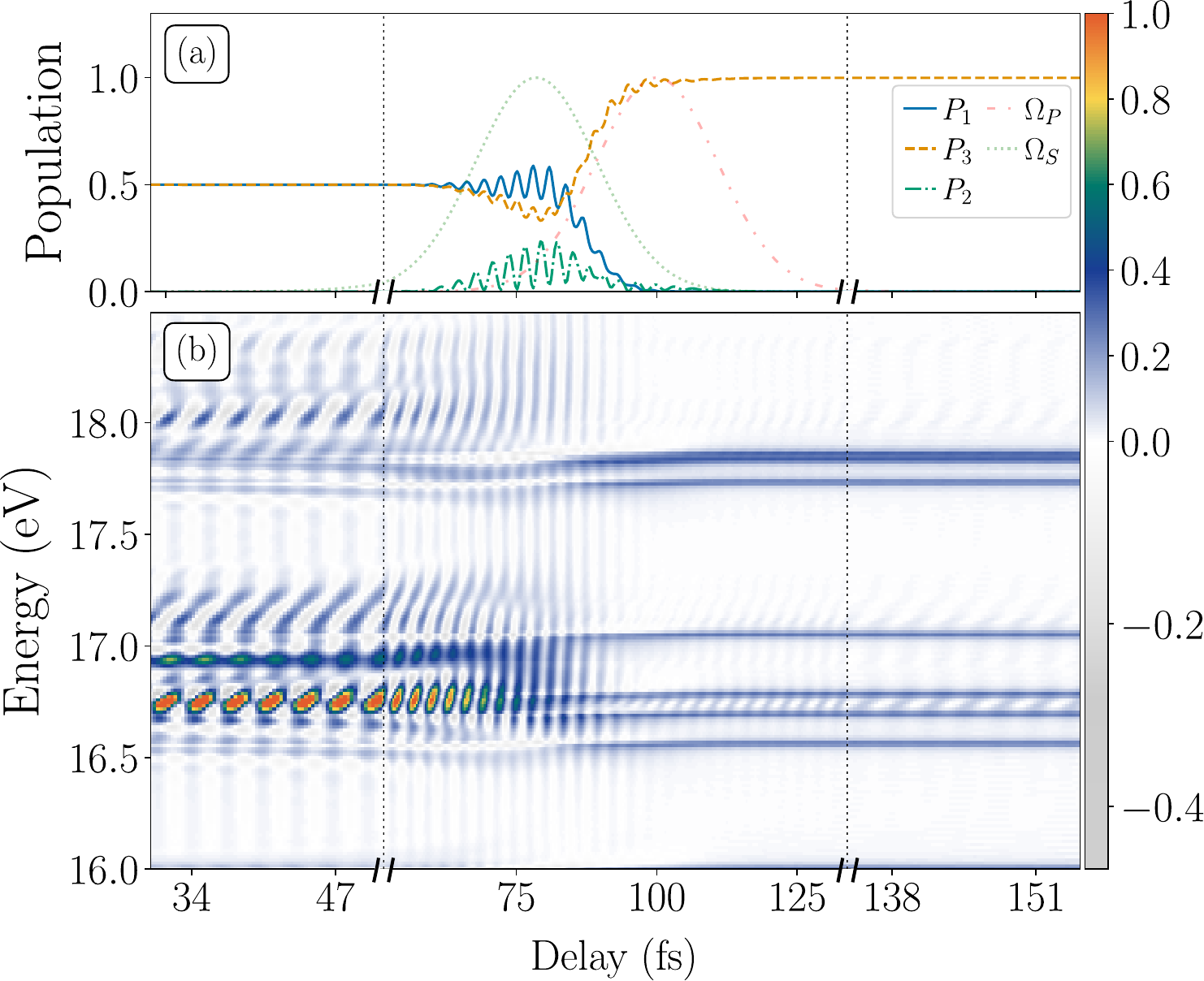}
    \caption{Control of electron motion in spin-orbit split $5s^2 5p^5 (^2P^{\circ}_{3/2})$ and $5s^2 5p^5 (^2P^{\circ}_{1/2})$ ground electronic states of the xenon cation via high-lying $4s5s^25p^6 (^{2}S_{1/2})$ core-excited electronic state using the two delayed X-ray pulses with identical Gaussian envelopes. (a) Applied laser fields and the resulting population transfer. (b) Time-resolved transient absorption spectrogram. It is seen that the system is driven from an equal coherent mixture of the spin-orbit split states to the $^2P^{\circ}_{1/2}$ state of the system. The initially created charge migration dynamics is suppressed by the control fields, as indicated by the disappearance of oscillatory features in the simulated absorption spectra.}
    \label{fig:Xray_ATAS}
\end{figure}

\subsection{X-ray control via core ionic state}
\label{subsec:Xray_control}
Let us now investigate possibilities to use x-ray pulses for the population transfer and charge migration control in Xe$^+$. Given that this setup is even closer to the three-state Hamiltonian in Eq.~(\ref{eq:hamil}), owing to the isolation of the intermediate state, all the schemes proposed above are capable of effectively controlling the wave packet dynamics. In order to illustrate this, we first present a scheme based on two pulses with identical Gaussian envelopes to achieve population transfer starting from an even superposition of the spin-orbit split ground states. Then, we demonstrate an application of the analytical generalized f-STIRAP scheme for enhancing the initially weak charge migration dynamics by driving the system from a coherent superposition of states $\ket 1$ (10\%) and $\ket 3$ (90\%) to an equal superposition of these states.

\begin{figure}
    \centering
    \includegraphics[width=\linewidth]{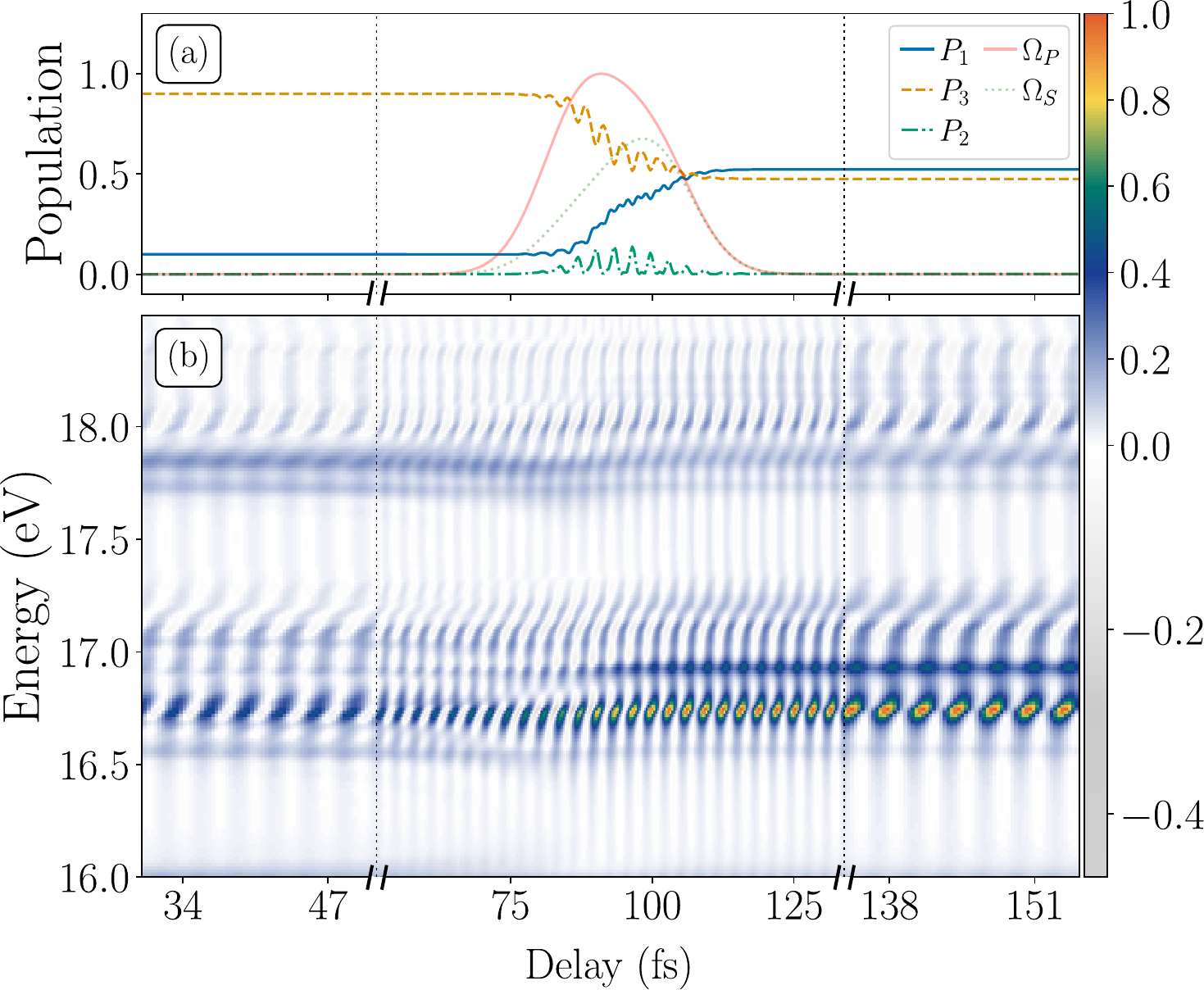}
    \caption{Control of electron motion in spin-orbit split $5s^2 5p^5 (^2P^{\circ}_{3/2})$ and $5s^2 5p^5 (^2P^{\circ}_{1/2})$ ground electronic states of xenon cation via high-lying $4s5s^25p^6 (^{2}S_{1/2})$ core-excited electronic state using the analytical generalized f-STIRAP method. Top panel: Applied laser fields and the resulting population transfer. Bottom panel: Time-resolved transient absorption spectrogram. It is seen that the initially weak oscillatory features in the spectra are significantly enhanced when the system is driven from a coherent superposition of states $^2P^{\circ}_{3/2}$ (10\%) and $^2P^{\circ}_{1/2}$ (90\%) to an equal superposition of these states.}
    \label{fig:Xray_ATAS_enhancement}
\end{figure}

In our simulations employing two pulses with identical Gaussian envelopes, we tuned the pump and Stokes pulses to be exactly on resonance with the transitions to state$\ket{2'}$. The utilized intensities and durations of the pulses are $E_P = E_S = 10.3$~TW/cm$^2$ and $\gamma_P = \gamma_S = 18$~fs, respectively. The phase offset and the delay between the pulses required to drive the system from an initial state composed of an equal superposition of states $\ket{1}$ and $\ket 3$ to a state purely composed of state $\ket 3$ are found to be $\phi = 1.25 \pi$ and $\Delta t = -21.4$~fs. For the probe step, we employ the ATAS scheme targeting the same energy region and using the same parameters as described in Sec.~\ref{subsec:XUV_control}. The results, presented in Figs.~\ref{fig:Xray_ATAS}(a) and~\ref{fig:Xray_ATAS}(b), show the successful population transfer, and therefore precise control over the charge migration dynamics, in the realistic Xe$^+$ system.

In Fig.~\ref{fig:Xray_ATAS_enhancement}, we show an implementation of the generalized f-STIRAP control scheme utilizing the pulse envelopes given by Eq.~(\ref{eq:two_ps}). Using $E_P=E_S=10.88$~TW/cm$^{2}$, $\gamma_L=\gamma_R=12$~fs, $\alpha\approx0.1\pi$, $\beta=\pi/4$, $t_L=87.56$~fs, and $t_R=100$~fs, we investigate the possibility to enhance the degree of coherence between the eigestates of the system by transferring population from an initial superposition of states $\ket 1$ (10\%) and $\ket 3$ (90\%) to their nearly equal superposition. The calculated absorption spectrum in Fig.~\ref{fig:Xray_ATAS_enhancement}(b) shows how the contrast of the oscillations is significantly increased as the coherence between states $\ket 1$ and $\ket 3$ is enhanced.

We note that the laser pulses required for x-ray STIRAP can be efficiently generated using modern free electron laser (FEL) facilities such as LCLS-II, SwissFEL, or the upcoming CXFEL at Arizona State University, to name just a few. These sources not only can produce intense ultrashort pulses with the frequencies required to target well isolated core states, but they are also capable of generating radiation of multiple colors~\cite{huang2021}. With regards to the latter, it has been shown that it is possible to produce two-color X-ray bursts with a high degree of tunability in their delay and bandwidth~\cite{hara2013,marinelli2015,ferrari2016}. Without going into the details, in order to generate this type of soft x-ray pulse it is possible to use a split undulator which induces the radiation at different frequencies, that with the addition of a chicane allows for the appropriate time separation. Alternatively, one can use more advanced electron bunch modification via seeding lasers that change the properties of the electron bunch and are capable of inducing the generation of two color x-rays with the added benefit of achieving low timing jitter since the radiation is produced by the same electron bunch.

\section{Conclusions}
\label{sec:conclusions}
In conclusion, this paper presents an extension of the STIRAP technique to scenarios where control is applied to an arbitrary linear superposition of two quantum states. We developed the formalism and working equations giving access to parameters of laser pulses which can drive the coherent dynamics of a $\Lambda$-type three-level system. We presented a generalization of the f-STIRAP that enables arbitrary control over the mixing ratio of quantum states within a wave packet.

We investigated the performance of an approximate variant of this generalized f-STIRAP approach, where the analytically designed laser pulses with composite envelopes are replaced by fitted Gaussians. We found that as long as the mixing angle of the adiabatic state varies as fast as it does in the analytical model between the initial and final values, the population transfer achieved with the fitted pulses is comparable to that taking place with the original fields. Importantly, we also found that the population dynamics is done preserving the exclusion of state $\ket 2$ from the mixture, with some minor oscillations being added on top of otherwise smooth adiabatic transition. 
 
As an alternative to adiabatic population transfer taking place in STIRAP techniques, we studied the performance of the control scheme with two pulses with identical Gaussian envelopes tuning their relative phase and delay. We found that if the intensity of the pulses is decreased, taking the system out of adiabaticity, it is possible to find values of delay and the phase such that practically any initial mixture of states can be transformed to a desired final state. In particular, we examined the case when the complete transfer to either states $\ket 1$ or $\ket 3$ is performed. We found that the regions of parameters landscapes for which this can be achieved vary smoothly, indicating that the scheme is robust to small fluctuations of the applied fields. Since in this case the system is taken out of adiabaticity, state $\ket 2$ gets transiently populated as the transfer occurs. Nonetheless, depending on the parameters, we found that the population change in one of the states is monotonic, which can be useful in certain quantum control applications. 

We conducted numerical investigations of the performance of our developed control schemes using the xenon cation as a model system. We explored the possibilities to control the ultrafast charge migration dynamics originating from a coherent superposition of the spin-orbit split $5s^2 5p^5 (^2P^{\circ}_{3/2})$ and $5s^2 5p^5 (^2P^{\circ}_{1/2})$ ground electronic states of the xenon ion. We employed two types of control pulses: those in the XUV range, utilizing the low-lying $5s5p^6 (^{2}S_{1/2})$ valence excited state, and those in the x-ray region, accessing the $4s5s^25p^6 (^{2}S_{1/2})$ core excited state. In both cases we were able to perform various types of population transfer, highlighting the efficiency of our developed approaches in different regimes. We demonstrated how the implemented control over the electron dynamics in the system can be observed using time-resolved absorption spectroscopy.

We complemented our study by briefly reviewing current experimental capabilities to generate XUV and x-ray pulses that can be used in the reported control schemes. We discussed perspectives of using HHG sources in combination with a TDCM in order to isolate harmonics of interest from the spectra and manipulate their time duration and intensity. We also reflected on the potential of using x-rays coming from FEL sources that are currently capable of generating two-color pulses in the soft x-ray regime with well-controlled delay and phase offset between them.

This work represents a proof-of-concept theoretical study, offering a foundation for future theoretical advancements and experimental investigations. We leave the discussion of possible extensions, such as the briefly mentioned use of different polarizations for each one of the pulses involved in the control scheme, for future research studies. We note that the presented control techniques can be directly implemented in more complex systems, such as Rydberg atoms or molecules where the possibilities of controlling charge migration dynamics attract significant interest from the community. One aspect to consider when applying these control schemes to molecules is the inherent loss of coherence in the electronic dynamics due to nuclear motion. It is, therefore, expected that the control of the populations of the involved electronic states should be implemented prior to the complete decoherence of the initiated electronic dynamics. Furthermore, the motion of the nuclear wave packets along the potential energy surfaces of the corresponding electronic states can alter energy separations between the initial and intermediate states, as well as the associated dipole couplings. This should not necessarily be a hurdle to achieve population control in these more complicated scenarios as there has been work, like the analytical treatment presented in Ref.~\cite{Carrol1988}, that shows control of such time dependent systems. We hope that our work will motivate further studies exploring applications of the generalized f-STIRAP for controlling quantum dynamics in realistic systems. %The main hurdle would be to find a combination of pulse sources and system that lends itself to a smooth implementation that also allows for measurable dynamics. 
 
\begin{acknowledgments}
The authors wish to express their sincere gratitude to Arvinder Sandhu for the excellent criticisms of the manuscript, and for innumerable discussions during the development of this work. The computational part of this research is based upon High Performance Computing (HPC) resources supported by the University of Arizona TRIF, UITS, and Research, Innovation, and Impact (RII) and maintained by the UArizona Research Technologies department. The authors acknowledge the financial support from the U.S. Department of Energy (DOE), Office of Science, Basic Energy Sciences (BES) under Award \#DE-SC0024182. %\cite{Einwohner1976Analytical,Carroll1987Three}
\end{acknowledgments}

\section*{Data availability}
The data that support the findings of this article are openly available~\cite{paper_data}.

\appendix

\begin{widetext}
\section{Computation of Dipole Matrix Elements Between Spin-Orbit Coupled Electronic States of Xenon Cation}
\label{sec:appendix_dip}
We seek to have an efficient way to calculate electric dipole matrix elements between spin-orbit coupled electronic states, including high-lying Rydberg states, of singly ionized xenon atom. To avoid the difficulties of \textit{ab initio} electronic structure calculations, we employ a semiempirical methodology by utilizing data from the NIST Atomic Spectra Database~\cite{NIST_ASD_2024}, which provides state energies and the corresponding electronic configurations of atomic states. We assume that the electronic configuration given in the database accurately represents the dominant configuration of a bound state and compute transition dipole matrix elements using the algorithm detailed below.

In singly ionized xenon, states are classified using both $LS$ and $JK$ angular momentum coupling schemes. The two lowest spin-orbit split states satisfy $LS$ coupling and the first excited state (those used in the main text as the three states of the STIRAP scheme) are dominantly $LS$ coupled, whereas all higher excited states follow $JK$ coupling. 

A typical $JK$ coupled state of xenon is specified by the following quantum numbers
\begin{equation}
5\text{s}^2 5\text{p}^4 (^{2S_p+1}L_{p J_p} ) n'L_E'
\quad
{}^{2S_E+1}[K]
\quad
J.
\end{equation}
The quantum number $K$ is formed by coupling $J_p$ and $L_E'$; $K$ is then coupled to the spin of the outer electron $S_E$ to form $J$. From this, we can calculate the dipole moments assuming that the dipole operator will only act on the orbital momentum of the outermost electron.

The procedure is different depending on the coupling between the two states. Between two $JK$ coupled states, we first apply the Wigner--Eckart theorem:
\begin{equation}
\bra{\gamma' j' m'}
T_k^q
\ket{ \gamma j m}
=
(-1)^{j'-m'}
\begin{pmatrix}
j' & k & j \\
-m' & q & m
\end{pmatrix}
\redmat{\gamma' j'}{T_k}{\gamma j}.
\end{equation}
To obtain the reduced matrix element acting on the single outermost electron we recursively apply Eq.(7.1.7) from Ref.~\cite{edmonds2016angular}
\begin{equation}
\redmat{\gamma' (j_1' j_2') J'}{T_k(1)}{\gamma(j_1 j_2) J }
=
(-1)^{j_1'+j_2+J+k}
\sqrt{ (2J+1) (2J'+1) }
\begin{Bmatrix}
j_1' & J' & j_2 \\
J & j_1 & k
\end{Bmatrix}
\redmat{\gamma' j_1'}{T_k(1)}{\gamma j_1}
\end{equation}
in order to decouple the angular momenta until the outermost electron's angular momentum is isolated. 

This results in:
\begin{equation}
\begin{aligned}
\redmat{\gamma' J' (S_E', K'(J_P'(L_P', S_P'), L_E') )}{ r Y_l}{\gamma J (S_E, K(J_P(L_P, S_P), L_E) )}
=
&
\\[5mm]
 (-1)^{K'+S_E+J+l}
[J][J']
\begin{Bmatrix}
K' & J' & S_E \\
J & K & l
\end{Bmatrix} \redmat{\gamma' K'(J_P'(L_P', S_P'), L_E')}{ r Y_l}{\gamma J K(J_P(L_P, S_P), L_E)}
=
&
\\[5mm]
(-1)^{\substack{L_E'+J_P+K+l+K'+S_E+J+l}} [K][K'][J][J']
\begin{Bmatrix}
L_E' & K' & J_P \\
K & L_E & l
\end{Bmatrix}
\begin{Bmatrix}
K' & J' & S_E \\
J & K & l
\end{Bmatrix} \redmat{\gamma' L_E'}{r Y_l}{\gamma L_E },
\\
\end{aligned}
\end{equation}
where $[x]=\sqrt{2x+1}$. To calculate $\redmat{\gamma' L_E'}{r Y_l}{\gamma L_E}$ in the outermost electron, the radial component of the wavefunction is assumed to be a hydrogen-like orbital with the principal quantum number taken from the reported electronic configuration, and the charge $Z=2$.\\

The angular part of the reduced matrix element can be computed analytically, and in terms of the radial integral the reduced matrix element is
\begin{equation}
\redmat{n'l'}{r Y_l }{n l}
=(-1)^{l'}[l'][l]\sqrt{\frac{3}{4 \pi} } \begin{pmatrix} l' & 1 & l \\ 0 & 0 & 0 \end{pmatrix}
\int R_{n,l}(r) \ r\ R_{n',l'}(r)\ r^2 dr.
\end{equation}

To calculate the dipole matrix elements between a $JK$ coupled state and an $LS$ coupled states we have three cases to deal with. The coupling with the state $5s5p^6 (^{2}S)_{1/2} $ is assumed to be zero since a transition to any of the $JK$ coupled states would require two electrons to change orbitals, which would be against the Slater-Condon rules. Therefore, in our model this state will be only coupled with the spin-orbit split ground states, which should be a good approximation. For the dipole matrix elements between $5\text{s}^2 5\text{p}^5 (^2P)J$ and all higher excited $JK$ states, we apply the Wigner-Eckart theorem but neglect the angular momentum coupling of innermost electrons since the ground state does not allow for a clear identification of the angular momentum of the core electrons. Therefore, we approximate the dipole moment by assuming that the reduced matrix element is that of a hydrogenlike transition between a $5p$ electron and the valence electron of the excited state.

\end{widetext}

\bibliography{references.bib}
%\bibliography{library.bib} %temporary bibliography file with the whole library

\end{document}